\documentclass[aps,prb,twocolumn,groupedaddress,superscriptaddress,nobibnotes]{revtex4-1}
\usepackage{amsmath}
\usepackage{amssymb}
\usepackage{graphicx}
\usepackage{dcolumn}
\usepackage{bm}
\usepackage{color}
\usepackage{subfig}
\usepackage[normalem]{ulem}

\newcommand{\MBT}{MnBi$_2$Te$_4$}
\newcommand{\MBTBT}{MnBi$_4$Te$_7$}
\newcommand{\MBTBTBT}{MnBi$_6$Te$_{10}$}

\newcommand{\BT}{Bi$_2$Te$_3$}

\newcommand{\PreserveBackslash}[1]{\let\temp=\\#1\let\\=\temp}

\begin{document}


\title{Variety of magnetic topological phases in the (MnBi$_2$Te$_4$)(Bi$_2$Te$_3$)$_m$ family }




\author{I.\,I.~Klimovskikh}
\email{ilya.klimovskikh@spbu.ru}
\affiliation{Saint Petersburg State University, 198504 Saint Petersburg, Russia}

\author{M.\,M. Otrokov}
\email{mikhail.otrokov@gmail.com}
\affiliation{Centro de F\'{i}sica de Materiales (CFM-MPC), Centro Mixto CSIC-UPV/EHU,  20018 Donostia-San Sebasti\'{a}n, Basque Country, Spain}
\affiliation{IKERBASQUE, Basque Foundation for Science, 48011 Bilbao, Spain}

\author{D.~Estyunin}
\affiliation{Saint Petersburg State University, 198504 Saint Petersburg, Russia}

\author{S.\,V.~Eremeev}
\affiliation{Institute of Strength Physics and Materials Science,
Russian Academy of Sciences, 634055 Tomsk, Russia}
\affiliation{Tomsk State University, 634050 Tomsk, Russia}
\affiliation{Saint Petersburg State University, 198504 Saint
Petersburg, Russia}

\author{S.\,O.~Filnov}
\affiliation{Saint Petersburg State University, 198504 Saint Petersburg, Russia}

\author{A.~Koroleva}
\affiliation{Saint Petersburg State University, 198504 Saint Petersburg, Russia}

\author{E.~Shevchenko}
\affiliation{Saint Petersburg State University, 198504 Saint Petersburg, Russia}

\author{V.\,~Voroshnin}
\affiliation{Helmholtz-Zentrum Berlin f\"ur Materialien und Energie,
Elektronenspeicherring BESSY II, Albert-Einstein-Straße 15, 12489 Berlin, Germany}

\author{I.\,P. Rusinov}
\affiliation{Tomsk State University, 634050 Tomsk, Russia}
\affiliation{Saint Petersburg State University, 198504 Saint Petersburg, Russia}

\author{M.\, Blanco-Rey}
\affiliation{Departamento de F\'{\i}sica de Materiales UPV/EHU, 20080 Donostia-San Sebasti\'{a}n, Basque Country, Spain}
\affiliation{Donostia International Physics Center (DIPC), 20018 Donostia-San Sebasti\'{a}n, Basque Country, Spain}

\author{M.\,~Hoffmann}
\affiliation{Institut f\"ur Theoretische Physik, Johannes Kepler Universit\"at, A 4040 Linz, Austria}

\author{Z.S.~Aliev}
\affiliation{Azerbaijan State Oil and Industry University, AZ1010 Baku, Azerbaijan}
\affiliation{Institute of Physics, National Academy of Sciences of Azerbaijan, AZ1143  Baku, Azerbaijan}

\author{M. B. Babanly}
\affiliation{Institute of Catalysis and Inorganic Chemistry, Azerbaijan National Academy of Science, AZ1143 Baku, Azerbaijan}
\affiliation{Baku State University, AZ1148 Baku, Azerbaijan}

\author{I. R. Amiraslanov}
\affiliation{Institute of Physics, National Academy of Sciences of Azerbaijan, AZ1143  Baku, Azerbaijan}
\affiliation{Baku State University, AZ1148 Baku, Azerbaijan}



\author{N.A. Abdullayev}
\affiliation{Institute of Physics, National Academy of Sciences of Azerbaijan, AZ1143  Baku, Azerbaijan}

\author{V.N. Zverev}
\affiliation{Institute of Solid State Physics RAS, Chernogolovka, Moscow district, Russia}

\author{A. Kimura}
\affiliation{Department of Physical Sciences, Graduate School of Science, Hiroshima University, 1-3-1 Kagamiyama, Higashi-Hiroshima 739-8526, Japan}

\author{O.E.\,Tereshchenko}
 \affiliation{A.V. Rzhanov Institute of Semiconductor Physics, Novosibirsk, 630090 Russia}
 \affiliation{Novosibirsk State University, Novosibirsk, 630090 Russia}
\affiliation{Saint Petersburg State University, 198504 Saint Petersburg, Russia}

   \author{K. A. Kokh}
\affiliation{V.S. Sobolev Institute of Geology and Mineralogy, Siberian Branch, Russian Academy of Sciences, Novosibirsk, 630090 Russian Federation}
\affiliation{Novosibirsk State University, Novosibirsk, 630090 Russia}
\affiliation{Saint Petersburg State University, 198504 Saint Petersburg, Russia}

\author{L. Petaccia}
\affiliation{Elettra Sincrotrone Trieste, Strada Statale 14 km 163.5, 34149 Trieste, Italy}

\author{G. Di Santo}
\affiliation{Elettra Sincrotrone Trieste, Strada Statale 14 km 163.5, 34149 Trieste, Italy}

\author{A.~Ernst}
\affiliation{Institut f\"ur Theoretische Physik, Johannes Kepler Universit\"at, A 4040 Linz, Austria}
\affiliation{Max-Planck-Institut f\"ur Mikrostrukturphysik, Weinberg 2, D-06120 Halle, Germany}

\author{P.\,M.~Echenique}
\affiliation{Donostia International Physics Center (DIPC), 20018 Donostia-San Sebasti\'{a}n, Basque Country, Spain}
\affiliation{Departamento de F\'{\i}sica de Materiales UPV/EHU, 20080 Donostia-San Sebasti\'{a}n, Basque Country, Spain}
\affiliation{Centro de F\'{i}sica de Materiales (CFM-MPC), Centro Mixto CSIC-UPV/EHU,  20018 Donostia-San Sebasti\'{a}n, Basque Country, Spain}

\author{N. T. Mamedov}
\affiliation{Institute of Physics, National Academy of Sciences of Azerbaijan, AZ1143  Baku, Azerbaijan}

\author{A.\,M.~Shikin}
\affiliation{Saint Petersburg State University, 198504 Saint Petersburg, Russia}

\author{E.\,V. Chulkov}
\email{evguenivladimirovich.tchoulkov@ehu.eus}
\affiliation{Centro de F\'{i}sica de Materiales (CFM-MPC), Centro Mixto CSIC-UPV/EHU,  20018 Donostia-San Sebasti\'{a}n, Basque Country, Spain}
\affiliation{Departamento de F\'{\i}sica de Materiales UPV/EHU, 20080 Donostia-San Sebasti\'{a}n, Basque Country, Spain}
\affiliation{Donostia International Physics Center (DIPC), 20018 Donostia-San Sebasti\'{a}n, Basque Country, Spain}
\affiliation{Saint Petersburg State University, 198504 Saint Petersburg, Russia}

\date{\today}

\begin{abstract}

Quantum states of matter combining non-trivial topology and magnetism attract a lot of attention nowadays; the special focus is on  magnetic topological insulators (MTIs) featuring quantum anomalous Hall and axion insulator phases.  
 Feasibility of many novel phenomena that \emph{intrinsic} magnetic TIs may host depends crucially on our ability to engineer and efficiently tune their electronic and magnetic structures. Here, using angle- and spin-resolved photoemission spectroscopy along with \emph{ab initio} calculations we report on a large family of intrinsic magnetic TIs in the homologous series of the van der Waals compounds (MnBi$_2$Te$_4$)(Bi$_2$Te$_3$)$_m$   with $m=0, ..., 6$. Magnetic, electronic and, consequently, topological properties of these materials depend strongly on the $m$ value and are thus highly tunable. The antiferromagnetic (AFM)   coupling  between the neighboring Mn layers strongly weakens on moving from MnBi2Te4 (m = 0) to MnBi4Te7 (m = 1), changes to ferromagnetic (FM) one in  MnBi6Te10 (m = 2) and  disappears  with further increase in m. 
In this way, the AFM and FM TI states are respectively realized in the $m=0,1$ and $m=2$ cases, while for $m \ge 3$ a novel and hitherto-unknown topologically-nontrivial phase arises, in which below the corresponding critical temperature the magnetizations of the non-interacting 2D ferromagnets, formed by the \MBT\, building blocks, are disordered along the third direction.
The variety of intrinsic magnetic TI phases in (MnBi$_2$Te$_4$)(Bi$_2$Te$_3$)$_m$ allows efficient engineering of functional van der Waals heterostructures for topological quantum computation, as well as antiferromagnetic and 2D spintronics.

\end{abstract}

\maketitle

\section{Introduction}

Magnetism and topology can meet each other both in real space,
giving rise to complex magnetic structures such as vortices or
skyrmions, and in reciprocal momentum space, resulting in Weyl
semimetal or magnetic topological insulator (MTI) phases. In the MTI
case, the interplay between topology and magnetism provides
particularly rich playground for realization of exotic physics.
Below the magnetic critical temperature, the time-reversal symmetry
breaks down introducing a mass term to the linear dispersion of
Dirac fermions thus opening opportunities for realization of such
phenomena as  quantized anomalous Hall (QAH) and magnetoelectric
effect, axion electrodynamics, or Majorana
fermions.\cite{Liu.prl2008, He2014, Chang.natm2015, Chang.sci2013,
Feng.prl2015} These unusual properties make MTIs extremely
attractive for applications in  novel electronics and in the
emerging 2D\cite{Gibertini2019, Burch2018, Lin2019aa} and
antiferromagnetic (AFM) spintronics. \cite{Jungwirth.nnano2016,
Smejkal.nphys2018, Baltz.rmp2018}

Until recently, the magnetism in TIs has been achieved using either
magnetic doping\cite{Chang.sci2013, Chang.natm2015, Checkelsky.natphys2012, Chen.sci2010,
Henk.prl2012, Hor.prb2010, PhysRevB.97.245407}, or proximity
effect\cite{Eremeev.prb2013, Katmis2016} as well as by construction of  van der Waals (vdW)
heterostructures \cite{Otrokov.jetpl2017, Otrokov.2dmat2017,
Hirahara.nl2017, Eremeev:2018aa}. This situation has changed
drastically with the recent discovery of the AFM TI
phase\cite{Mong.prb2010} in stoichiometric vdW layered
antiferromagnet MnBi$_2$Te$_4$ \cite{Otrokov.arxiv2018,
Otrokov.prl2019, Zhang.prl2019, Li.sciadv2019, Gong.cpl2019}, which
inspired a lot of research activity as it holds promise for
realization of the high-temperature QAH and axion insulator states,
Majorana hinge modes and other effects \cite{Mong.prb2010,
Otrokov.prl2019, Zhang.prl2019, Li.sciadv2019, Peng.prb2019,
Deng.arxiv2019, Liu_arxiv2019, Zhang2019, Zhang.arxiv2019a}.

\begin{figure*}[!bth]

\includegraphics[width=1.0\textwidth]{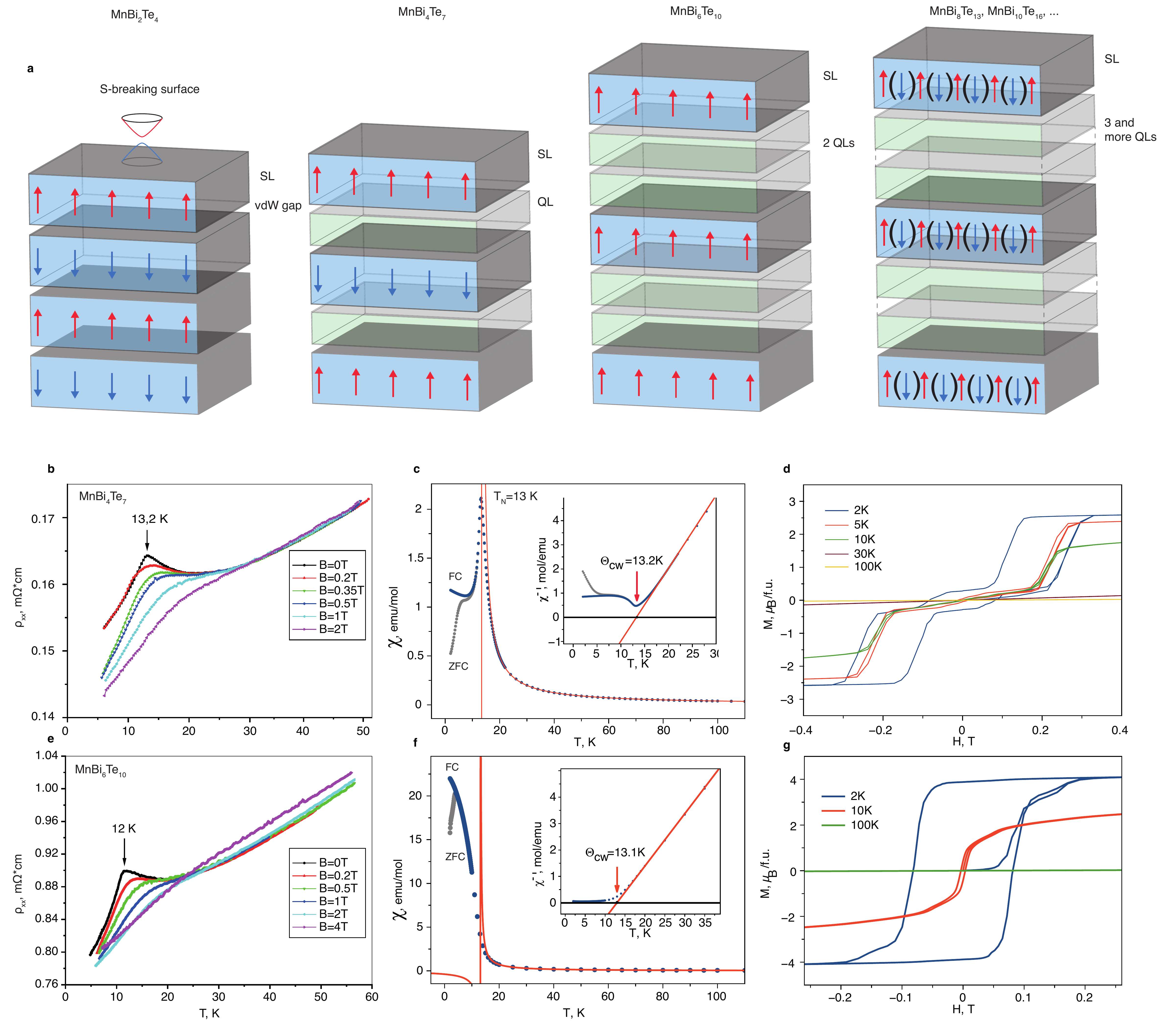}

\caption{\bf Magnetic structure of  (MnBi$_2$Te$_4$)(Bi$_2$Te$_3$)$_m$ series.}

{\bf a} Compounds  (MnBi$_2$Te$_4$)(Bi$_2$Te$_3$)$_m$ consist of alternating  five-layer (QL) and magnetic seven-layer (SL) blocks    {\bf b,e} Temperature dependence of the resistivity for MnBi$_4$Te$_7$ and MnBi$_6$Te$_{10}$ for various applied out-of-plane magnetic fields. {\bf c,f} Magnetic susceptibility for MnBi$_4$Te$_7$ (c) and MnBi$_6$Te$_{10}$ (f) as a function of temperature measured in an external magnetic field of 0.1\,T in zero-field-cooled and field-cooled conditions. Temperature-dependent reciprocal susceptibility is shown in the corresponding insets.The red line is a  Curie-Weiss fit to the high-temperature data (see text for details). {\bf d,g} Field dependent  magnetization curves taken at various temperatures with out-of-plane external magnetic field for MnBi$_4$Te$_7$ (d) and MnBi$_6$Te$_{10}$ (g).

\label{Fig1}
\end{figure*}

Here we  propose the MTI family by
introducing the (MnBi$_2$Te$_4$)(Bi$_2$Te$_3$)$_m$ series of vdW materials that,
apart from \MBT\, ($m=0$), contains six more
topologically-nontrivial compounds, namely MnBi$_4$Te$_7$ ($m=1$),
MnBi$_6$Te$_{10}$ ($m=2$), MnBi$_8$Te$_{13}$ ($m=3$),
MnBi$_{10}$Te$_{16}$ ($m=4$), MnBi$_{12}$Te$_{19}$ ($m=5$) and
MnBi$_{14}$Te$_{22}$ ($m=6$).   Along the series, the strength of
the interlayer exchange interaction, that couples neighboring FM Mn
layers, gradually decreases with the increase of $m$, while its
character changes from an AFM ($m=0,1$) to FM ($m=2$). It is then
followed by a crossover into the purely 2D magnetic regime starting
from $m=3$. Combined with  the non-trivial topology of the
(MnBi$_2$Te$_4$)(Bi$_2$Te$_3$)$_m$ compounds, these magnetic states
give rise to the AFM and FM TI phases for $m=0,1$ and $m=2$,
respectively, while for $m\geq3$ a new
MTI phase is formed in which, below the respective $T_\text{C}$, the magnetizations
of the 2D FM-ordered Mn layers of the \MBT\, building blocks are
disordered along the [0001] directions. The topologically nontrivial
nature of  these compounds is confirmed by ARPES measurements
that reveal the presence of the topological surface (TSS) states
whose dispersion depends strongly on the crystal surface
termination. The unusual magnetic properties  make the
(MnBi$_2$Te$_4$)(Bi$_2$Te$_3$)$_m$ series a unique tunable platform
for creating various exotic states of matter such as intrinsic axion
or QAH insulator\cite{Otrokov.2dmat2017, Otrokov.prl2019,
Liu_arxiv2019}, the field induced QAH insulator\cite{Deng.arxiv2019,
Liu_arxiv2019} or chiral topological superconductor.\cite{Zhang2019}

\section{Crystal  structure and magnetic crossover}

The first compound in the (MnBi$_2$Te$_4$)(Bi$_2$Te$_3$)$_m$ series is MnBi$_2$Te$_4$  (m=0), which had been investigated previously and discovered  to be the  first AFM TI\cite{Otrokov.arxiv2018}. This system consists of septuple layer (SL) blocks stacked one on top of another.
Each SL  is  a 2D ferromagnet, while the coupling between the neighboring SLs is antiferromagnetic.\cite{Otrokov.arxiv2018}
For $m \geq 1$, the members of the (MnBi$_2$Te$_4$)(Bi$_2$Te$_3$)$_m$ family are comprised of alternating septuple (\MBT) and quintuple (\BT) layer  blocks,
see Fig.\ref{Fig1}a. The growth details and crystal structures of thus-formed MnBi$_4$Te$_7$
and MnBi$_6$Te$_{10}$ compounds have been first reported in  Ref.\onlinecite{Aliev.jac2019}, for which the -5-7-5-7- and -5-5-7-5-5-7- blocks sequences corresponding to $m=1$ and $m=2$  have been respectively revealed.

The XRD patterns of our MnBi$_4$Te$_7$, 
MnBi$_6$Te$_{10}$, MnBi$_8$Te$_{13}$, MnBi$_{10}$Te$_{16}$,
MnBi$_{12}$Te$_{19}$, and MnBi$_{14}$Te$_{22}$ samples, shown in
Fig.~S1 of Supplementary Information, confirm their $P\bar 3m1$ and
$R\bar 3m$ space groups.

\begin{figure*}
\includegraphics[width=1\textwidth]{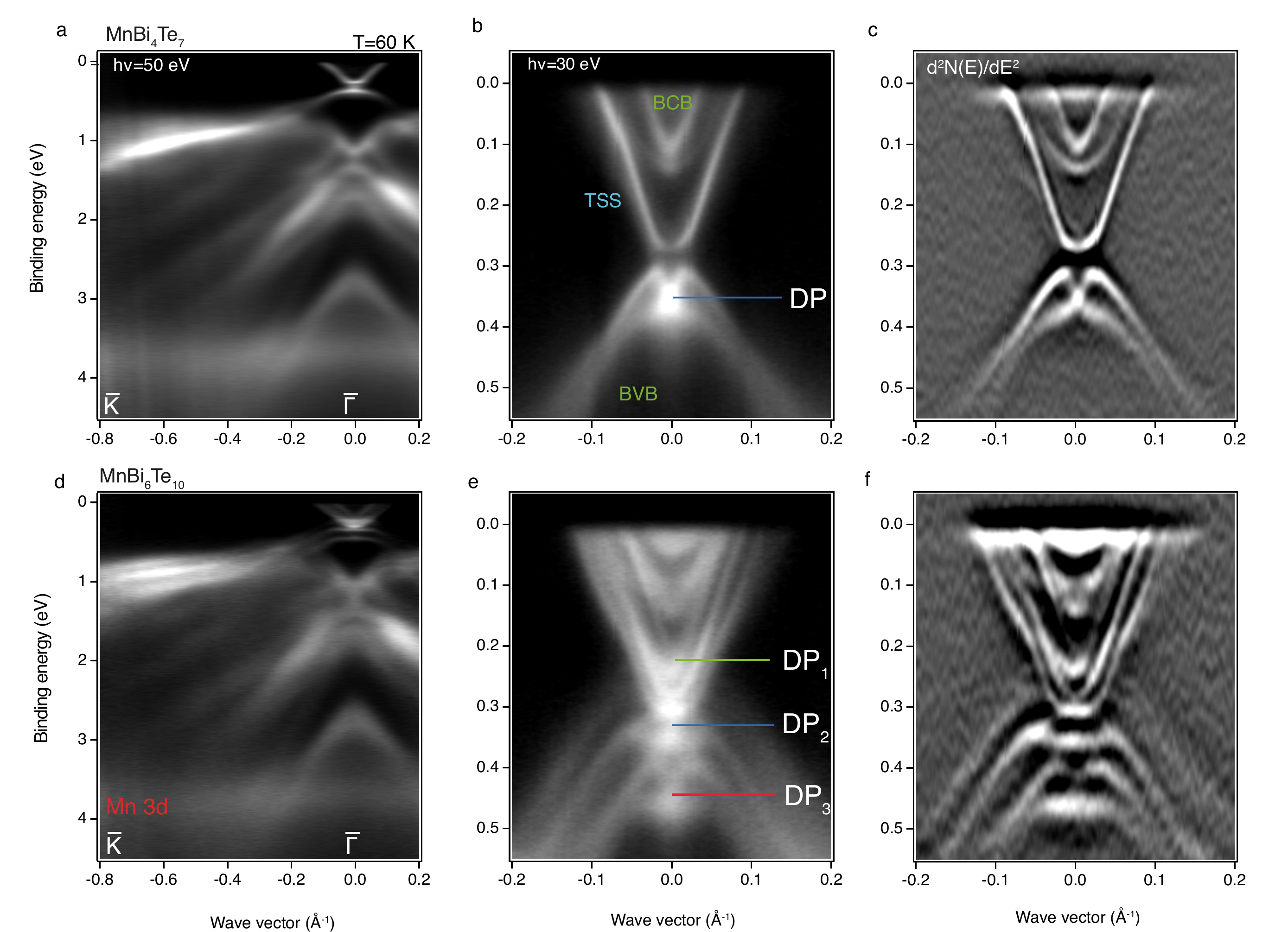}
\caption{{\bf ARPES measurements of  MnBi$_4$Te$_7$ and MnBi$_6$Te$_{10}$.}
{\bf a,d}, ARPES dispersion relations for the large energy and momentum region of MnBi$_4$Te$_7$ and MnBi$_6$Te$_{10}$ taken at  $60\,\mathrm{K}$ at a photon energy of $50\,\mathrm{eV}$; {\bf b,e},   ARPES dispersion relations of the Dirac cone region taken at a photon energy of $30\,\mathrm{eV}$ at $60\,\mathrm{K}$; {\bf c,f}  Second derivatives of the spectra presented in b,e.
}
 \label{Fig2}
\end{figure*}

The temperature dependence of the resistivity measured in a zero magnetic field  demonstrates the metallic-like behaviour for both MnBi$_4$Te$_7$ and MnBi$_6$Te$_{10}$, see Fig.\ref{Fig1}b,e. At low temperatures, the well-defined kinks  at around $12-13\,\mathrm{K}$ are observed for both compounds.  As the external magnetic field is switched on and increased these features are washed out, pointing towards their magnetic origin. However, while thus-determined critical temperatures are  almost the same for MnBi$_4$Te$_7$ and MnBi$_6$Te$_{10}$, the magnetic orders are different for these two compounds, as shown by means of SQUID magnetometry.

The field-cooled (FC) and zero-field cooled (ZFC) magnetic susceptibilities measured at $0.1\,\mathrm{T}$ for MnBi$_4$Te$_7$ are shown in Fig.\ref{Fig1}c. The high temperature paramagnetic behaviour ends up with a pronounced peak at $13\,\mathrm{K}$, that is characteristic of an AFM ordering. The fitting of the susceptibility by the Curie-Weiss law yields the Curie-Weiss temperature ($\Theta_{CW}$) of $13.2\,\mathrm{K}$. Positive sign and relatively high value of $\Theta_{CW}$(c.f. $3-6\,\mathrm{K}$ for MnBi$_2$Te$_4$ \cite{Otrokov.arxiv2018}) indicate the presence of strong ferromagnetic interactions in MnBi$_4$Te$_7$ in spite of the overall AFM behaviour.
The N\'eel temperature of only $13\,\mathrm{K}$ for \MBTBT\, versus $25\,\mathrm{K}$ for MnBi$_2$Te$_4$ reveals a strong weakening of the interlayer AFM coupling due to the insertion of the non-magnetic quintuple layer block (QL) between neighboring SLs. The onset of the ferromagnetic interactions is also  manifested in splitting of the FC and ZFC curves at low temperature (spin-glass-like transition).

The $M(H)$ curves taken at various temperatures are presented in Fig.\ref{Fig1}d. Slightly below the N\'eel temperature ($10\,\mathrm{K}$) a typical AFM $M(H)$ behaviour takes place, with a spin-flip upturns appearing at $0.2\,\mathrm{T}$, the latter value being much lower than that for MnBi$_2$Te$_4$ ($3.5\,\mathrm{T}$)\cite{Otrokov.arxiv2018}. At very low temperatures, the spin-flip hysteresis opens up in the magnetic field interval from 0.1 to $0.3\,\mathrm{T}$, as seen for $T=2\,\mathrm{K}$. Moreover, in the absence of  external field, the magnetization does not fall to zero and forms a ferromagnetic hysteresis loop from -0.1  to $0.1\,\mathrm{T}$. Such a dual complex metamagnetic behaviour can be explained by the presence of domains with  ferro- and antiferromagnetic ordering between neighbor SLs. The presented data testify the weak 3D AFM/FM and strong 2D FM intralayer coupling in MnBi$_4$Te$_7$. Similar results with the metamagnetic behavior of MnBi$_4$Te$_7$ have been reported recently in Refs.\onlinecite{ Sun.arxiv2019, Hu.arxiv2019, Wu.arxiv2019,   Vidal.arxiv2019-2}

The described experimental picture is consistent with that yielded by theory.
Namely, the DFT exchange coupling parameters calculations reveal, first, a stable tendency to the intralayer FM ordering in the SLs of \MBTBT\, and, second, a strong drop of the interlayer exchange coupling (1-2 orders of magnitude) as compared to \MBT\, (see Supplementary Note II).
Highly-accurate total-energy calculations give an energy difference between the FM and interlayer AFM configurations of $0.25\,\mathrm{meV}$ per Mn pair in favor of the interlayer AFM one. The magnetic anisotropy energy, $E_a$, is positive and equals to $0.12\,\mathrm{meV}$ (i.e. the easy axis is out-of-plane), which is roughly two times smaller than in bulk \MBT. In good agreement with the resistivity and magnetization measurements, our Monte Carlo simulations yield a N\'eel temperature of $13.6\,\mathrm{K}$ for the bulk \MBTBT. The drop of the N\'eel temperature from about $25\,\mathrm{K}$ in \MBT\, to $\simeq$$13\,\mathrm{K}$ in \MBTBT\, is precisely caused by the weakening of the interlayer exchange coupling.

\begin{figure*}
\includegraphics[width=0.9\textwidth]{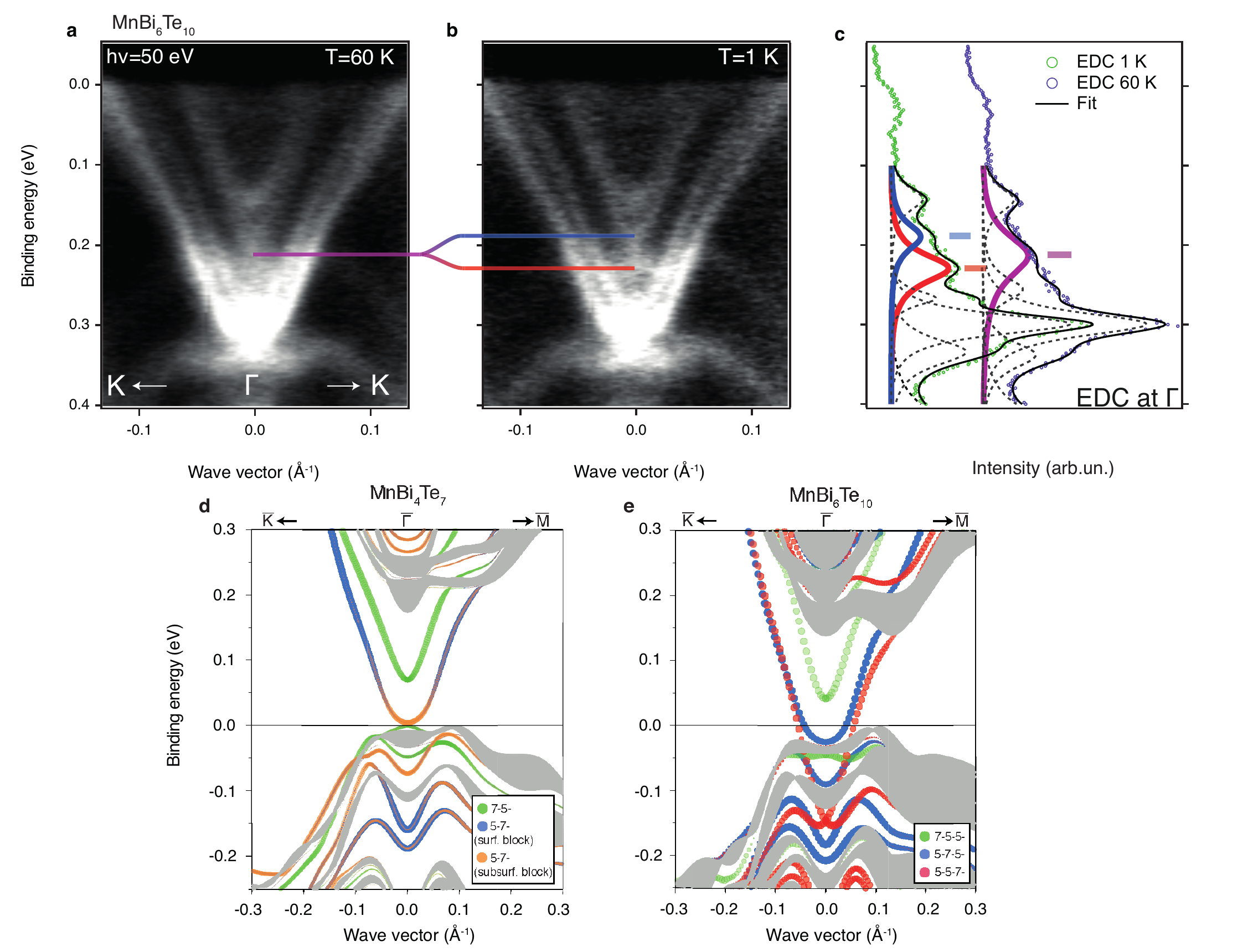}
\caption{{\bf Temperature dependent ARPES measurements  and DFT calculations}
{\bf a-c} ARPES measurements for MnBi$_6$Te$_{10}$ taken at 60 and $1\,\mathrm{K}$, respectively and corresponding EDC profiles at $\overline{\Gamma}$-point with fitting results. The Dirac point gap opening can be seen as splitting of the magenta peak at $60\,\mathrm{K}$ to red and blue peaks at $1\,\mathrm{K}$.
    {\bf d}, Layer-resolved electronic structure of the MnBi$_4$Te$_7$ (0001) surface. The size of color circles reflects the weight of the state in the topmost block of the SL (green) and QL (blue) termination. The orange circles show the contribution of the subsurface (SL) block of the QL termination. The grey areas correspond to the bulk bandstructure projected onto the surface Brillouin zone. {\bf e}, The same for MnBi$_6$Te$_{10}$ (0001) surface, the size of green, blue, and red circles reflects the weight of the state in the topmost block of the 7-5-5-, 5-7-5-, and 5-5-7- terminations.
}
 \label{Fig3}
\end{figure*}

Increasing by one the number of QLs between SLs and forming MnBi$_6$Te$_{10}$  not only leads to  further weakening of the interlayer exchange interaction, but also changes the character of the latter. The magnetic susceptibility data for this compound, shown in Fig.\ref{Fig1}f, do not exhibit any peak related to AFM ordering, while the FC and ZFC curves splitting and the Curie-Weiss fit ($\Theta_{CW}=13.1\,\mathrm{K}$) reveal the same dominating intralayer ferromagnetic interactions as for  MnBi$_4$Te$_7$. A pronounced hysteresis showing no spin-flip transitions is observed in the $M(H)$ curves, demonstrating an almost pure ferromagnetic behavior. Some deviations of the $M(H)$ curve slope at $0.05\,\mathrm{T}$ can be related to  residual antiferromagnetic coupled domains.
The DFT total-energy calculations performed for \MBTBTBT\, show that the interlayer coupling weakens with respect to the \MBTBT\, case, the difference between the AFM and FM states being $0.1\,\mathrm{meV}$ in favor of the AFM ordering. The discrepancy with the experimental results indicates that an overall FM behaviour may be caused by the $n$-doped character of the samples. 

 The $m=3$ and 4 members of the series, i.e.
MnBi$_8$Te$_{13}$ and MnBi$_{10}$Te$_{16}$, show an overall
ferromagnetic behavior below the Curie temperatures of about
$12\,\mathrm{K}$, as revealed by the magnetization measurements (see
Supplementary Fig.~S4); a similar behaviour is expected for
MnBi$_{12}$Te$_{19}$ ($m=5$) and MnBi$_{14}$Te$_{22}$ ($m=6$)
because of the vanishing interlayer exchange interaction. In Ref.
[\onlinecite{Otrokov.prl2019}], the Curie temperature of the
free-standing \MBT\, SL has theoretically been  predicted to be
$12\,\mathrm{K}$, which coincides with the values of
MnBi$_8$Te$_{13}$ and MnBi$_{10}$Te$_{16}$.  These facts allow us to
conclude that the interlayer exchange coupling between SLs is
practically absent in (MnBi$_2$Te$_4$)(Bi$_2$Te$_3$)$_m$ with $m>2$
and their  magnetic structure  can be described as a set of
decoupled 2D ferromagnets whose magnetizations are randomly oriented
-- either parallel or antiparallel to the [0001] direction. Note
that owing to the $n$-doping of the samples, the long-range RKKY
exchange interaction may still weakly couple the magnetic layer
blocks even for $m>2$, although no indication of that has been
observed in the experiment. Thus, the interlayer exchange
interaction between the 2D FM SLs in the
(MnBi$_2$Te$_4$)(Bi$_2$Te$_3$)$_m$ series first changes from AFM for
$m=0$ and 1 (\MBT\, and \MBTBT) to FM for $m=2$ (MnBi$_6$Te$_{10}$)
and then practically disappears for $m \ge 3$
(MnBi$_8$Te$_{13}$,  MnBi$_{10}$Te$_{16}$,
MnBi$_{12}$Te$_{19}$ and MnBi$_{14}$Te$_{22}$). As we show next,
these crossovers have profound consequences for the topological
properties of the materials in this series.

\section{Topological surface states}

We now turn to the electronic structure study of the compounds in
the (MnBi$_2$Te$_4$)(Bi$_2$Te$_3$)$_m$ series. As it has been
mentioned above, the MnBi$_2$Te$_4$ ($m=0$) compound was shown to be
an AFM TI, its (0001) surface exhibiting the TSS with a gapped Dirac
cone-like dispersion even above the T$_N$. The ARPES dispersion relations for the (0001)
surface of the MnBi$_4$Te$_7$ compound ($m=1$) measured at
$T=60\,\mathrm{K}$ (low temperature measurements will be  discussed below) are shown in Fig.~\ref{Fig2}a. Within the binding
energy (BE) window from 0.5 to $4\,\mathrm{eV}$, a quite complex
bandstructure, characteristic of bismuth chalcogenides is seen
\cite{Shikin2014, Bansil2016}. However, unlike  nonmagnetic TIs, a
weakly dispersing state is discernible in Fig.~\ref{Fig2}a at the BE
of about $3.8\,\mathrm{eV}$, which is attributed to the Mn
3$d$-states, whose intensity is enhanced at the Mn-resonant photon
energy of $50\,\mathrm{eV}$. Detailed ARPES dispersion relations in
the low energy part of the spectrum are presented in
Fig.~\ref{Fig2}b ($h\nu=30\,\mathrm{eV}$). Around the
$\overline{\Gamma}$-point, there are two electron pockets that can be
attributed to the bulk conduction bands of MnBi$_4$Te$_7$, clearly
visible in the second derivative of the spectra, presented in
Fig.~\ref{Fig2}c. Thus the material is $n$-doped which is typical
of Bi-containing TIs \cite{Hasan2010}. The third electronic band,
crossing the Fermi level, has largely a linear dispersion except for
the close vicinity of the $\overline{\Gamma}$-point, where it becomes
practically flat at the BE of $0.27\,\mathrm{eV}$. At slightly
higher BE, there is a M-shaped state, which is accompanied by a
hole-like band with which it is degenerate at the Brillouin zone
(BZ) center. Thus, between the linear and M-shaped bands a local gap
is formed. Below the gap the electron-like part of the latter band
behaves like a continuation of the linear band, as both have
practically the same group velocity. To separate the bulk and
surface states, ARPES measurements with various photon energies (50,
18  and  $6\,\mathrm{eV}$ (laser)) have been performed, as presented
in Supplementary Figure S6. The dispersion of the linear band does
not show any photon energy dependence, which points towards its
surface or 2D character. This allows  to suggest that the linear
band is the TSS, while far from the $\overline{\Gamma}$-poin the M-shaped band is likely to be the
trivial surface state in the valence band  that interacts
with the TSS near the $\overline{\Gamma}$-point, leading to the appearance of the
``avoided crossing'' gap at the BE of $0.29\,\mathrm{eV}$. In this
case, the degeneracy point between the M-band and the close-lying
hole band can be identified as the Dirac point (DP) of the
MnBi$_4$Te$_7$(0001) TSS. Since the TSS should be spin-polarized,
spin-ARPES measurements have been performed. The acquired spectra
(Supplementary Figure S6) clearly demonstrate the  spin polarization
reversal for opposite branches of the linear band, revealing the
characteristic helical spin texture of the TSS. The performed bands
assignment  allows us to estimate the bulk band gap of
MnBi$_4$Te$_7$ to be about $0.15\,\mathrm{eV}$. It should be
noted that on the surface of MnBi$_4$Te$_7$, since it is built of two
different crystal blocks, QL and SL, two surface terminations and,
consequently, two Dirac TSSs with different dispersion are expected
in the ARPES spectrum. The absence of the second TSS will be
discussed below.

The TSSs are also observed for the next member in
(MnBi$_2$Te$_4$)(Bi$_2$Te$_3$)$_m$ series,  MnBi$_6$Te$_{10}$
($m=2$). The ARPES dispersion relations for this
compound are shown in Fig. \ref{Fig2}d, measured at the same
conditions as the spectrum in Fig. \ref{Fig2}a. In the BE region
from 1 to $5\,\mathrm{eV}$, the valence band dispersions for
MnBi$_6$Te$_{10}$ look  similar to those of MnBi$_4$Te$_{7}$.
Apparently, the intensity of the Mn-3$d$ resonant feature seen at
the BE of 3.8 eV is slightly decreased  for
MnBi$_6$Te$_{10}$, in accordance with the lower relative Mn
concentration. The most pronounced difference between the band
structures of the two compounds takes place in the region of  BE
between $1\,\mathrm{eV}$ and E$_F$, where the surface states
dominate the photoemission signal. In particular, in Fig.
\ref{Fig2}d one can see three weakly dispersing states along the
$\overline{\Gamma} - \overline{\rm K}$ direction around the BE of $1\,\mathrm{eV}$ as well as
three Dirac cones at the BZ center. This complex dispersion of the
TSSs is resolved in the detailed ARPES image and its second
derivative, shown in Fig.\ref{Fig2}e and f, where both the lower and
upper parts of the three Dirac cones can be clearly seen.

The appearance of three TSSs at the MnBi$_6$Te$_{10}$(0001) surface
is caused by the peculiar crystal structure of the ($m=2$) compound,
built of the two types of blocks,  stacked  on top of each other in
the -5-5-7-5-5-7- sequence. In such a case, the cleavage procedure
leads to the formation of different surface terminations. Such a
structure was previously observed for the PbBi$_6$Te$_{10}$(0001)
surface, where the three terraces were attributed to the 5-5-7-,
5-7-5- and 7-5-5- terminations, showing different dispersions of the
TSS. \cite{Papagno.acsn2016} Similarly to PbBi$_6$Te$_{10}$, the
cleaved surface of MnBi$_6$Te$_{10}$ may exhibit three types of
terminations, two of them having a QL on the surface (i.e. either
5-5-7- or 5-7-5-), while the third possible crystal truncation is
that by the SL (7-5-5-). Similarly to the PbBi$_6$Te$_{10}$ case,
these terraces have  different Dirac cone binding energies and
spatial depth localization. The 
inner Dirac state (which by analogy with nonmagnetic TIs with
similar structure can be assigned to the TSS generated by the
SL-terminated terraces) shows the lowest BE
($\sim$$0.25\,\mathrm{eV}$) of the Dirac point (DP1) which is
located above the bulk valence band states, while the other
two cones (coming from different QL terraces) overlap with the bulk valence band (BVB).

The bulk phase transition from a paramagnetic to a ferromagnetic
state is expected to affect the surface electronic structure.
According to the DFT calculations shown  in Fig.~\ref{Fig3}e, in the FM state the inner Dirac
cone  exhibits a giant mass gap of
about $90\,\mathrm{eV}$. To reveal such a behaviour in the experiment,
we have measured ARPES spectra for MnBi$_6$Te$_{10}$ above and below
the critical temperature, as shown in Fig.~\ref{Fig3}a,b. At
$60\,\mathrm{K}$, the DP of the SL termination TSS, located at a BE
of $0.22\,\mathrm{eV}$, consists of one broadened peak, as can be
proved by energy distribution curves (EDCs) analysis presented in
Fig.~\ref{Fig3}c. When  the sample is cooled down to $T =
1\,\mathrm{K}$, the Dirac point splits by about $50\,\mathrm{meV}$:
the EDC fitting clearly shows the opening of the gap at the Dirac
point. Note that this observation is possible because the DP
of the SL termination (DP1 in Fig ~\ref{Fig2}e) is well separated
from the BVB, which is not the case for the QL terminations of
 MnBi$_6$Te$_{10}$. Strikingly, the DP
exchange gap revealed for the SL-terminated MnBi$_6$Te$_{10}$(0001)
surface is closed in the paramagnetic phase, contrary to what
happens in MnBi$_2$Te$_{4}$. The reason of this difference can
lie in the reduced anisotropic spin fluctuations in comparison with
MnBi$_2$Te$_{4}$\cite{Otrokov.arxiv2018}, owing to the  larger SL-SL
distance and suppressed antiferromagnetic interaction. Finally,
the electronic structure measurements performed for the  $m=3$
MnBi$_8$Te$_{13}$(0001) and $m=4$ MnBi$_{10}$Te$_{16}$(0001)
surfaces reveal the presence of the TSSs within the bulk band gap
too (see Supplementary Information), but their analysis is
complicated due to increased number of surface terraces after the
cleavage.

 According to the DFT electronic structure calculations,  AFM bulk MnBi$_4$Te$_7$ features a fundamental band gap of $0.18\,\mathrm{eV}$ (Supplementary Note II). Because of its interlayer AFM ordering, \MBTBT\, is invariant with respect to the combination of the time-reversal ($\Theta$) and primitive-lattice translation ($T_{1/2}$) symmetries, $S=\Theta T_{1/2}$, and thus obeys the $Z_2$ topological classification of AFM insulators\cite{Mong.prb2010,Fang.prb2013}. We find $Z_2=1$ for \MBTBT, 
meaning that its fundamental band gap is inverted whereby, similarly to MnBi$_2$Te$_4$, \MBTBT\, is an AFM TI below the N\'eel temperature. The AFM TI state of \MBTBT\, along with the out-of-plane direction of the staggered magnetization's easy axis dictate that there should be a gapped (gapless) TSS at the $S$-breaking ($S$-preserving) surface. This is exactly what we find:
the TSS is gapped at both of the possible terminations of the MnBi$_4$Te$_7$(0001) surface, which is $S$-breaking. At the SL termination, the DP gap is located inside the fundamental bulk band gap and reaches a large value of $70\,\mathrm{meV}$. At the QL termination, a more complex surface electronic structure is revealed, with four bands located in the region of interest. 
To understand this spectrum, we have performed surface electronic
structure calculations at different spin-orbit coupling (SOC) strengths. As it can be seen
in Supplementary Fig.~S9, with no SOC included there is an
exchange-split trivial surface state in the valence bands' projected
band gap around the $\overline{\Gamma}$-point at energy of about
-0.3 eV. When the SOC is turned on and the system is in the
topological phase, this spin-split state interacts with the TSS of
the QL-termination, leading to the appearance of  avoided crossings
(Supplementary Fig.~S9).
At the natural SOC strength (Fig.~\ref{Fig3}d), these avoided crossing effects are quite significant and the Dirac cone appears to be torn in two parts, one of which is located in the fundamental band gap, while another in the projected band gap between approximately -0.1 and -0.2 eV. The split DP of the QL termination lies at about $0.16\,\mathrm{eV}$ below the Fermi level. 
The smaller DP gap size at the QL termination (about $29\,\mathrm{meV}$) is caused by the predominant localization of the TSS in the surface block, whereby the interaction of this state with the Mn layer, lying in the subsurface block, is weaker. Thus, the complex dispersion of the QL-termination TSS is caused by the interaction with the trivial surface state, located in the bulk valence band.
A comparison between the experimental and theoretical spectra allows
identifying the measured surface bandstructure as that of the QL
termination. In the PM state, the trivial surface state is
exchange-unsplit which leads to a less complex appearance of the
overall spectrum: there is only one avoided crossing clearly seen in
ARPES in Fig. \ref{Fig2}b. Thus, the gap in the TSS at a BE of about
$0.27\,\mathrm{eV}$ (Fig. \ref{Fig2}b) is indeed related to the
``cut'' of the Dirac cone by a trivial surface state and therefore
this gap is observed above the N\'eel temperature. Cooling down
below the magnetic critical point does not lead to significant
changes (see Supplementary Figure~S7), except for a slight
broadening of the BVB, most likely related to the exchange splitting
of the bulk states. Noteworthy, the SL--teminated Dirac cone is not seen in the ARPES spectra, the
absence of its photoemission signal has also been reported for Pb-
and Ge-based TIs in Refs. \onlinecite{Eremeev:2012aa,
Muff_PhysRevB}. Finally, our tight-binding calculations of the
$S$-preserving ($10 \bar 1 1$) surface electronic structure reveal a
gapless Dirac cone, as expected for an AFM TI (see Supplementary
Figure~S3). 

The DFT-calculated (0001) surface band structure of the FM MnBi$_6$Te$_{10}$ is presented in Fig. \ref{Fig3}e, where three Dirac cones are clearly seen in agreement with the ARPES data.
Both the 7-5-5- and 5-7-5- terminations show essentially similar behaviors to those revealed for the \MBTBT(0001). The TSS of the 5-5-7-termination is mostly located in the surface QL block and therefore its DP is almost unsplit.

Thus, all the compounds of the presented  (MnBi$_2$Te$_4$)(Bi$_2$Te$_3$)$_m$ family are magnetic TIs, whose topological class changes with $m$. While MnBi$_2$Te$_4$ and MnBi$_4$Te$_7$ ($m=0,1$) are 3D AFM TIs, MnBi$_6$Te$_{10}$ ($m=2$) turns out to be a 3D FM TI because of the change of the interlayer exchange coupling character. Starting from $m=3$ the interlayer exchange interaction disappears, which has a very interesting consequence: MnBi$_8$Te$_{13}$, MnBi$_{10}$Te$_{16}$, MnBi$_{12}$Te$_{19}$, and MnBi$_{14}$Te$_{22}$  are the first examples of stoichiometric 3D MTI compounds in which the magnetizations of the 2D FM-ordered layers are disordered along the [0001] direction  below the corresponding critical temperature.

The peculiar magnetic properties of the
(MnBi$_2$Te$_4$)(Bi$_2$Te$_3$)$_m$ family combined with the
nontrivial topology of its constituents enable observation of
interesting effects for all of the presented members of the series.
Those effects would take advantage of the interlayer exchange
coupling tunability along the series, which is feasible through the
changing of the number of QLs separating the SLs. This allows one to
tailor both the strength and character of the interlayer exchange
coupling and even to switch it off starting from $m \geq 3$. One
particular consequence of it is that the magnetic structure, and
therefore the topological class of the compounds can be tuned by an
external magnetic field. This property is especially attractive in
the 2D limit, where the [MnBi$_2$Te$_4$]$_{1
\text{SL}}$/(Bi$_2$Te$_3$)$_{m \text{QL(s)}}$/[MnBi$_2$Te$_4$]$_{1
\text{SL}}$ sandwiches, that can be obtained by careful exfoliation
of the thin flakes from the single crystal surface,
\cite{Deng.arxiv2019, Liu_arxiv2019} turn to  either the intrinsic
zero plateau QAH state (also known as the axion insulator state) or
the Chern insulator state.  For $m \geq 3$, the latter state can be
achieved in zero external magnetic field as has been
 earlier predicted by
theory,\cite{Otrokov.2dmat2017} while for MnBi$_2$Te$_4$ and,
probably, for MnBi$_4$Te$_7$ an external magnetic field is needed to
achieve the quantized Hall effect. \cite{Deng.arxiv2019,
Liu_arxiv2019} However, in the  MnBi$_4$Te$_7$ case, the
strength of the critical field needed to overcome the AFM interlayer
exchange coupling should be much smaller than that used for
MnBi$_2$Te$_4$. The latter fact has also been pointed out as an
advantage for a possible realization of the topological
superconductor state  based on the
(MnBi$_2$Te$_4$)(Bi$_2$Te$_3$)$_m$ family, hosting the exotic
Majorana fermions \cite{Zhang2019}. From this point of view, the
(MnBi$_2$Te$_4$)(Bi$_2$Te$_3$)$_m$ series represents a unique and
highly tunable topological van-der-Waals platform for creation of
both exotic topological phases and functional devices for
antiferromagnetic and 2D spintronics as well as for topological
quantum computing.

\section{Conclusions}

In summary, we have reported the magnetic topological insulators
family (MnBi$_2$Te$_4$)(Bi$_2$Te$_3$)$_m$ consisting of Mn-based
magnetic septuple layer blocks separated by different number $m$ of
non-magnetic quintuple layers.  The interlayer exchange coupling
between the neighboring septuple layers can be tuned by changing $m$
giving rise to a crossover from the interlayer antiferromagnetic
ordering for $m=0,1$ (MnBi$_2$Te$_4$,  MnBi$_4$Te$_7$ ) to the
ferromagnetic one for $m=2$ (MnBi$_6$Te$_{10}$) and, finally, to the
complete disappearance of the interlayer interaction for $m \ge
3$ (MnBi$_8$Te$_{13}$, MnBi$_{10}$Te$_{16}$, MnBi$_{12}$Te$_{19}$
and MnBi$_{14}$Te$_{22}$). Combined with a non-trivial topology of
the (MnBi$_2$Te$_4$)(Bi$_2$Te$_3$)$_m$ compounds, proven by ARPES,
these magnetic states give rise to the AFM and FM TI phases for
$m=0,1$ and $m=2$, respectively, while for 
for $m \ge 3$ a new MTI phase is formed in which, below
$T_\text{C}$, the magnetizations of the 2D FM-ordered Mn layers of
the \MBT\, building blocks are disordered along the [0001]
direction. Depending on the surface terminations, a complex bundle
of the Dirac cones is resolved by means of DFT and ARPES, and the
magnetic gap in the Dirac point is found below the critical
temperature. The tunable magnetic and topological phases in
(MnBi$_2$Te$_4$)(Bi$_2$Te$_3$)$_m$ series allow engineering the
promising platforms not only for QAH effect, axion insulators, and
Majorana fermions, but also for emerging fields of antiferromagnetic
and van-der-Waals 2D spintronics.

\section*{Methods}
\subsection*{Electronic structure and total-energy calculations}

Electronic structure calculations were carried out within the density
functional theory using the projector augmented-wave (PAW) method
\cite{Blochl.prb1994} as implemented in the VASP code \cite{vasp1,
    vasp2}. The exchange-correlation energy was treated using the
generalized gradient approximation \cite{Perdew.prl1996}. The
Hamiltonian contained scalar relativistic corrections and the
SOC was taken into account by the second variation
method \cite{Koelling.jpc1977}. In order to describe the van der Waals interactions we made use of the DFT-D2 \cite{Grimme.jcc2006} and the DFT-D3 \cite{Grimme.jcp2010, Grimme.jcc2011} approaches, which gave similar results. The energy cutoff for the plane-wave expansion was set to $270\,\mathrm{eV}$. All structural optimizations were performed using a  conjugate-gradient algorithm and a force tolerance criterion for
convergence of $0.01\,\mathrm{eV}$/{\AA}. Spin-orbit coupling was always included
when performing relaxations. Depending on the particular task and geometry,
different 
grids for the Brillouin zones (BZs) sampling were used
(see below), all of them being $\overline \Gamma$-centered.

The Mn $3d$-states were treated employing the GGA$+U$ approach
\cite{Anisimov1991} within the Dudarev scheme
\cite{Dudarev.prb1998}. The $U_\text{eff}=U-J$ value for the Mn 3$d$-states
was chosen to be equal to $5.34\,\mathrm{eV}$, as in previous works on \MBT\, \cite{Otrokov.jetpl2017, Otrokov.2dmat2017, Eremeev.jac2017, Hirahara.nl2017, Otrokov.prl2019, Otrokov.arxiv2018}.

The bulk magnetic ordering was studied using total-energy calculations, performed for the FM and two different AFM states. Namely, we considered an interlayer AFM state and a noncollinear (NCL) intralayer AFM state, in which three spin sublattices form angles of 120$^\circ$ with respect to each other\cite{Eremeev.jac2017}. To model the FM and interlayer AFM structures in \MBTBT\, and \MBTBTBT, we used cells with 24 and 34 atoms, respectively. For the \MBTBT\, system, these calculations were performed with the respective 3D BZs sampled by the $25\times 25\times 5$ $k$-point grids. For the \MBTBTBT\, compound, which exhibits extremely weak interlayer exchange coupling,
the interlayer ordering was carefully studied using very
fine $k$-meshes up to $23\times 23\times 19$ points.
These meshes were also used in the magnetic anisotropy energy calculations.
For \MBTBT, the noncollinear
intralayer AFM configuration was treated using hexagonal bulk cell containing three
atoms per layer [$(\sqrt{3}\times\sqrt{3})R30^\circ$ in-plane periodicity] and a
$11\times 11\times 3$ BZ sampling. For \MBTBTBT, NCL configuration has not been considered since all the available experimental and theoretical data in the literature indicate that each \MBT\, SL block orders ferromagnetically, irrespectively of its structural environment.

The \MBTBT\, and \MBTBTBT\, semi-infinite surfaces were simulated
within a model of repeating films separated by a vacuum gap of a minimum of $10\,\mathrm{\AA}$.
The interlayer distances were optimized for the utmost SL or QL block of each
surface. Both the structural optimizations and static electronic structure calculations were performed
using a $k$-point grid of $11\times 11\times 1$ in the two-dimensional BZ.

The magnetic anisotropy energies, $E_\text{a}=E_\text{diff}+E_\text{d}$, were calculated taking into account both
the total-energy differences of various magnetization directions
$E_\text{diff}=E_\text{in-plane} - E_\text{out-of-plane}$, and the energy
of the classical dipole-dipole interaction, $E_\text{d}$. To calculate $E_\text{diff}$, the energies for three inequivalent
magnetization directions [cartesian $x$, $y$ (in-plane) and $z$ (out-of-plane)] were
calculated and $E_\text{diff}$ was determined as the difference $E_{\mathrm{in-plane}}$ -- $E_z$,
where $E_{\mathrm{in-plane}}$ is the energy of the most energetically favorable in-plane
direction of the magnetization. 
The total energies
were calculated self-consistently for all considered directions.
The energy convergence criterion was set to  $10^{-7}\,\mathrm{eV}$
providing a well-converged $E_\text{diff}$ (up to a few tenth of meV) and
excluding "accidental" convergence. A cutoff radius of at least
$20\,\mathrm{\mu m}$. was used to calculate $E_\text{d}$.

\subsection*{Exchange coupling constants calculations}
For the equilibrium structure of MnBi$_4$Te$_7$ obtained with VASP,
we calculated the Heisenberg exchange coupling constants
$J_{0,i}$ also from first principles, within the full-potential linearized augmented plane waves (FLAPW)
formalism \cite{bib:wimmer81} as implemented in \textsc{Fleur} \cite{bib:fleur}.
In the $J_{0,i}$ calculations SOC was neglected. We took the GGA+$U$ approach \cite{bib:anisimov97,bib:shick99}
under the fully localised limit  \cite{bib:anisimov93}, using similar settings as those in Refs.\onlinecite{Otrokov.arxiv2018, Otrokov.prl2019}
The self-consistent FLAPW basis set Monkhorst-Pack $k$-point sampling
of the first BZ and a cutoff of 3.4\,Hartree.
The cutoff energy for the density and potential expansions was 10.4\,Hartree.
Muffin tin sphere radii values of 2.74\,a.u. for Mn and 2.81\,a.u.
for Bi and Te atoms were used, and the partial wave functions were
expanded up to cutoffs of $l=8$. Mn, Bi, and Te contribute $4s3d$, $5s5p$,
and $6s6p$ valence electrons, respectively.

The $J_{0,i}$ constants were extracted by Fourier inversion of spin spirals in the reciprocal cell characterized by the $q$ vectors of a $19\times19\times3$ grid. These dispersion energies, calculated in the force
theorem approach,
converged below 0.1\,meV and allowed to add up to 150 neighbouring atoms
to the Fourier analysis.

\subsection*{ARPES measurements}

The experiments were carried out at  1-cubed UE-112 beamline at BESSY II in Berlin,  BaDElPh beamline\cite{Petaccia} at Elettra synchrotron in Trieste (Italy), BL-1, BL-9 and laser ARPES endstation at HiSOR in Hiroshima, and at Research Resource Center of Saint Petersburg State University
``Physical methods of surface investigation''  with a Scienta R4000 or a Specs Phoibos 150 energy analyzer. Samples were cleaved $in situ$ at the base pressure of  $6\times 10^{-11}\,\mathrm{mbar}$ . The crystalline order and cleanliness of the surface were verified by low energy electron diffraction (LEED) and X-ray photoelectron spectroscopy (XPS).

Spin-resolved ARPES measurements were performed at BL-9 beamline of HiSOR synchrotron in Hiroshima and APE beamline at Elettra synchrotron in Trieste (Italy).   The spectra were measured using VLEED spin detector. Total energy and angular resolutions were  $20\,\mathrm{meV}$ and 1.5$^\circ$, respectively.

\subsection*{SQUID magnetometry}

Magnetic measurements were carried out in the resource center ``Center for Diagnostics of Materials  for Medicine, Pharmacology and Nanoelectronics'' of the  SPbU Science Park using a SQUID magnetometer with a helium cryostat manufactured by Quantum Design. The measurements were carried out in a pull mode in terms of temperature and magnetic field. The applied magnetic field was perpendicular to the (0001) sample surface.

\subsection*{Resistivity measurements}

Resistivity measurements were done with a standard four-probe ac technique using a low-frequency (f$\propto$20 Hz) Lock-in amplifier. Contacts were attached with conducting graphite paste. The measurements were carried out in a temperature-variable cryostat at different values of magnetic field up to $8\,\mathrm{T}$, generated by a superconducting solenoid and directed along the normal to the (0001) sample surface.

\section*{Acknowledgments}
This work is supported by Saint Petersburg State
University grant for scientific investigations (No. 15.61.202.2015) and Russian Science Foundation ( grant No.
18-12-00062 in part of the photoemission measurements and 18-12-00169 in part of calculations of topological invariants, investigation of dependence the electronic spectra on SOC strength, and tight-binding band structure calculations).
Russian Foundation for Basic Research (Grant No. 18-52-06009) and
Science Development Foundation under the President of the Republic
of Azerbaijan (Grant No. EIF-BGM-4-RFTF-1/2017-21/04/1-M-02) are
acknowledged.  We also acknowledge the support by the Basque
Departamento de Educacion, UPV/EHU (Grant No. IT-756-13), Spanish
Ministerio de Economia y Competitividad (MINECO Grant No.
FIS2016-75862-P), and Tomsk State University competitiveness
improvement programme (project No. 8.1.01.2017).
 The calculations were performed in Donostia International Physics
Center and in the Research park of St. Petersburg State University Computing Center (http://cc.spbu.ru). \\

\section*{Authors contributions}
The manuscript has been written by I.I.K. and M.M.O. The ARPES measurements were carried out by I.I.K, A.M.S., D.E., V.V., S.O.F., G.S., L.P., A.Kimura. The SQUID magnetometry has been done by A.Koroleva. and E.S. The samples have been grown and characterized by Z.S.A., M.B.B., I.R.A, O.E.T. and K.A.K. The transport measurements have been carried out by N.Y.M., V.N.Z. and N.A.A. The band structure calculations were performed by M.M.O., S.V.E., I.P.R., M.B.R. The exchange coupling constants calculations were performed by M.B.R.,  M.M.O., and A.E. The magnetic anisotropy studies were performed by M.M.O. The Monte Carlo simulations were performed by M.H.  The supervision of the project was executed by E.V.C. All authors contributed to the discussion and manuscript editing. \\

\section*{Competing financial interests}
Authors declare no competing financial interests.

\bibliography{./paper}


\end{document}